\documentstyle[aps, psfig]{revtex}

\begin{document}
\title{An example of quantum control via Feshbach resonance in Bose-Einstein
condensates}
\author{S. Choi and N. P. Bigelow}
\address{Department of Physics and Astronomy, University of Rochester, Rochester, NY 
14627}
\maketitle

\begin{abstract}
We present a simple example of quantum control in Bose-Einstein condensates
via Feshbach resonance. By tuning an initially positive scattering length 
to zero, it is possible to generate oscillatory motion of the condensate 
that results from
quantum interference. The density oscillation is accompanied by a periodic
enhancement of the quantum mechanical squeezing of the amplitude quadrature.
\end{abstract}


\vspace{6mm}

With the rapid technological advances in quantum physics and nanoscience, it
is becoming increasingly necessary to develop methods of quantum control
based on the laws of quantum mechanics. Owing to the microscopic scale involved, one of the main avenues 
so far for the manipulation and control
has been the interaction of the quantum system with an electromagnetic wave.  Prominent examples include
atom optics and cavity QED\cite{atomoptics}.

The Bose-Einstein condensate (BEC), created experimentally almost a decade ago,
is a unique mesoscopic quantum system which is being studied intensely by a
great number of researchers from various disciplines. Many experimental
results have been reported in quick succession since the initial production
of BEC\cite{varenna}. One of the important, early BEC experiments
was that of inducing Feshbach resonances to tune the $s$-wave scattering
length\cite{original,theory1,theory2,MIT,Texas,JILA,Stanford}.
Recent experiments have provided the opportunity to tune the scattering length from the repulsive ($a > 0$) through the ideal gas ($a \sim 0$) to the attractive ($a <0$) regime.
Feshbach resonance has become an indispensible tool in the studies of BEC ranging from the production of molecular BEC\cite{molecule1,molecule2} 
to the observation of a ``bosenova'', a phenomenon closely related to the supernova\cite{bosenova,collapse1,cornish}. A large number of theoretical studies on the collapse and the subsequent ``explosion'' of BEC has since also appeared in the literature\cite{collapse2}.

A Feshbach resonance occurs when the energy of a pair of atoms in the condensate matches that of a metastable molecular state.  
The $s$-wave scattering length, which is a function of the energy difference between the atomic and the molecular states, may be varied near a
Feshbach resonance using magnetic field: 
\begin{equation}
a(B) = a_{b} \left ( 1 - \frac{\Delta B}{B -B_0} \right ),
\end{equation}
where, for the case of $^{85}$Rb in the $f = 2$, $m_f = -2$ hyperfine state,
a Feshbach resonance characterized by the width $\Delta B \approx 11.6$G and
the off-resonant background scattering length $a_b \approx -450a_0$ where $%
a_0$ is the Bohr radius, is found at $B_0 \approx 154.9$G
\cite{cornish}. We show in this paper that, in theory, tuning the scattering
length of BEC to zero via Feshbach resonace (e.g. by setting the magnetic
field to $B = \Delta B + B_0$) constitutes a simple method of quantum control -- observable changes are induced in the density of the BEC, and in addition, quantum mechanical squeezing\cite{walls} is realized. 

A recent, related theoretical work has considered periodic sign-changing modulation of the scattering length  similar to dispersion management schemes in fiber optics\cite{FRM}. This method was found to induce ``breathers'' which oscillate between the Thomas-Fermi (TF) and Gaussian configurations. Our proposed scheme, although similar, does not require changes in the sign of the scattering length. We propose a situation in which an initially positive scattering length $a>0$ is switched to $\epsilon$ where $\epsilon$ is negligibly small but positive. Because $a$ is at all times non-negative, this precludes the possibility of instability and collapse introduced when $a<0$\cite{collapse1,collapse2}. We study the subsequent evolution of the condensate as $a$ is kept constant at $\epsilon \approx 0$, away from the resonance.

As in Ref. \cite{FRM}, we assume that atom losses are minimized by careful experimental control\cite{JILA,molecule1,cornish}, and that we are in an off-resonant regime in which  
any molecules formed are rapidly destroyed, overdamping any atom-molecule population oscillations. In such a regime, the formation of a molecular condensate is suppressed and 
an accurate description of the condensate may be assumed to be provided by the Gross-Pitaevskii Equation (GPE). We write the time-independent GPE as follows
\begin{equation}
\left [ -\frac{\hbar^2}{2M} \nabla^{2} + V({\bf r}) + g |\psi_{c}({\bf r})
|^{2} \right ] \psi_{c}({\bf r}) = \mu \psi_{c}({\bf r}) ,
\end{equation}
where $\mu$ is the chemical potential, $V({\bf r})$ is the harmonic confining potential of the form $\frac{1}{%
2}M \omega_{t}^2 {\bf r}^2$, and $g = 4 \pi N a \hbar^2 /M$ is the interparticle coupling constant with the $s$-wave scattering length $a$ and the total number of atoms $N$. $\psi_{c}$ represents the condensate wave function which is normalized to one, $\int |\psi_{c}({\bf r})|^2 d {\bf r} = 1$, as we have included the total number of condensate atoms in the coupling constant.

\begin{figure}[t]
\centerline{\psfig{height=11cm,file=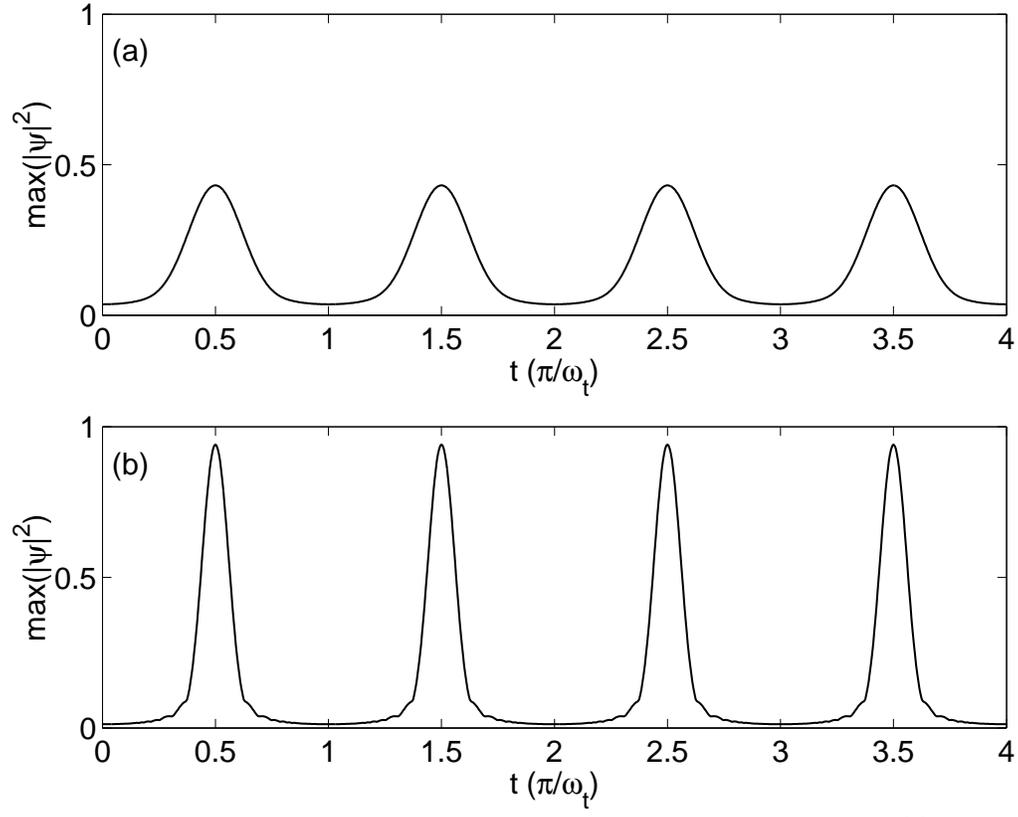}}
\caption{Maximum density variation of BEC after tuning the $s$-wave scattering length to zero. (a) for a small condensate; (b) for a BEC within the TF approximation. Time is given in units of $\pi/\omega_t$ where $\omega_t$ is the trap frequency.}
\label{density}
\end{figure}

If the scattering length $a$ is  set equal to zero suddenly  (within a time scale $t \ll 1/\omega_t$, to avoid any adiabatic following), the eigenstates of
the system become the harmonic oscillator states, $\phi_n({\bf r})$, 
with the corresponding eigenvalues $n \hbar \omega_t$. 
It is straightforward to see that the equation obeyed by $\psi_c({\bf r})$ after the switch to $a \sim 0$ can be written as a set of equations
\begin{equation}
i\hbar \frac{d \alpha_n}{d t} = n \hbar \omega_{t} \alpha_n , \label{alpha}
\end{equation}
where $\alpha_n \equiv \int \phi^{*}_{n} ({\bf r})
\psi_c ({\bf r}) d {\bf r}$ are the expansion coefficients of the BEC ground state expanded using a complete set of eigenfunctions $\{ \phi_n({\bf r}) \}$, $n = 0,1, \ldots$.
The coefficients $\alpha_n$ satisfy the normalization relation $\sum_n |\alpha_n|^2 = 1$, and may arbitrarily be chosen to be real.
The solution to Eq. (\ref{alpha}), which assumes that the switch to $a \sim 0$ occurs at time $t = 0$, is trivially given by $\alpha_n = \alpha_n(0) \exp(-i n \omega_{t} t)$. This causes the condensate wave function to become time-dependent: $\psi_c({\bf %
r}, t) = \sum_{n} \alpha_n(0) \exp(-i n \omega_{t} t) \phi_n ({\bf r})$. In
particular, the density of the BEC ground state is now given as 
\begin{eqnarray}
|\psi_c ({\bf r}, t) |^2 & = & \sum_{m,n} \alpha^{*}_m(0)\alpha_n(0) \exp [ -i
(n - m) \omega_{t} t ] \phi^{*}_m({\bf r}) \phi_n ({\bf r}) \\
& = & \sum_{n} |\alpha_n(0)|^2 | \phi_n({\bf r}) |^2 + \sum_{n > m} 2
\alpha_m(0)\alpha_n(0) \cos [ (n - m) \omega_{t} t ] \phi_m ({\bf r}) \phi_n
({\bf r}),  \label{thirdline}
\end{eqnarray}
where we have used the fact that $\alpha_n(0)$ and $\phi_n({\bf r})$ are
real. Equation (\ref{thirdline}) clearly demonstates a time-dependent
density oscillation resulting from quantum interference. The form
of the density is suggestive of a diffraction grating in time, in which
various components of the condensate wave function interfere with one
another. The ``visibility'' ${\cal V}$, of the
temporal interference at the center of the trap may be calculated by defining the ``intensity'' $I =
|\psi_c(0,t)|^2$: 
\begin{equation}
{\cal V} = \frac{I_{max} - I_{min}}{I_{max} + I_{min}} = \frac{ 2
\sum_{n > m} \alpha_m(0)\alpha_n(0) \phi_m (0) \phi_n (0)}{
\sum_{n} |\alpha_n(0)|^2 | \phi_n(0) |^2 } .
\end{equation}
The interference of various contributing components implies that the initial profile of the condensate is an important factor in determining the character of the density oscillation and the associated visibility. For a small BEC far from the TF limit, $\psi_c$ may well be approximated by a Gaussian. This implies that amongst all the expansion coefficients, $\alpha_0(0)$ and $\alpha_2(0)$ should be dominant. The second lowest coefficient, $\alpha_1(0)$, that corresponds to the first excited state $\phi_1({\bf r})$ is negligible since the condensate in a harmonic trap is an even function. Assuming all the other coefficients are very small,  we get
$|\psi ({\bf r}, t) |^2 \approx |\alpha_{0}(0)|^2 | \phi_{0} ({\bf r}) |^2 + |\alpha_{2}(0)|^2 | \phi_{2} ({\bf r}) |^2  + 2 \alpha_{0}(0)\alpha_{2}(0) \cos ( 2 \omega_{t} t )  \phi_{0}({\bf r}) \phi_{2} ({\bf r})$. Thus the condensate follows an oscillation at  frequency $2\omega_t$ with the corresponding visibility at the center of the trap given as ${\cal V} = {2 \alpha_0(0)\alpha_2(0) \phi_0(0) \phi_2 (0) }/ 
[ {\ |\alpha_0(0)|^2 | \phi_n (0) |^2 + |\alpha_2(0)|^2 | \phi_2(0) |^2 } ]$. 
Figure \ref{density} shows the maximum density variation of the BEC after tuning the 
scattering length to zero, with the top figure giving the result for a small condensate and the bottom figure for a condensate in the TF approximation.
As expected, different oscillation patterns are observed whereas the frequency of oscillation is $2\omega_t$ in both cases. This implies that in the TF approximation, the lower coefficients are still dominant, although with different weightings from that for a small condensate.

\begin{figure}[t]
\centerline{\psfig{height=11cm,file=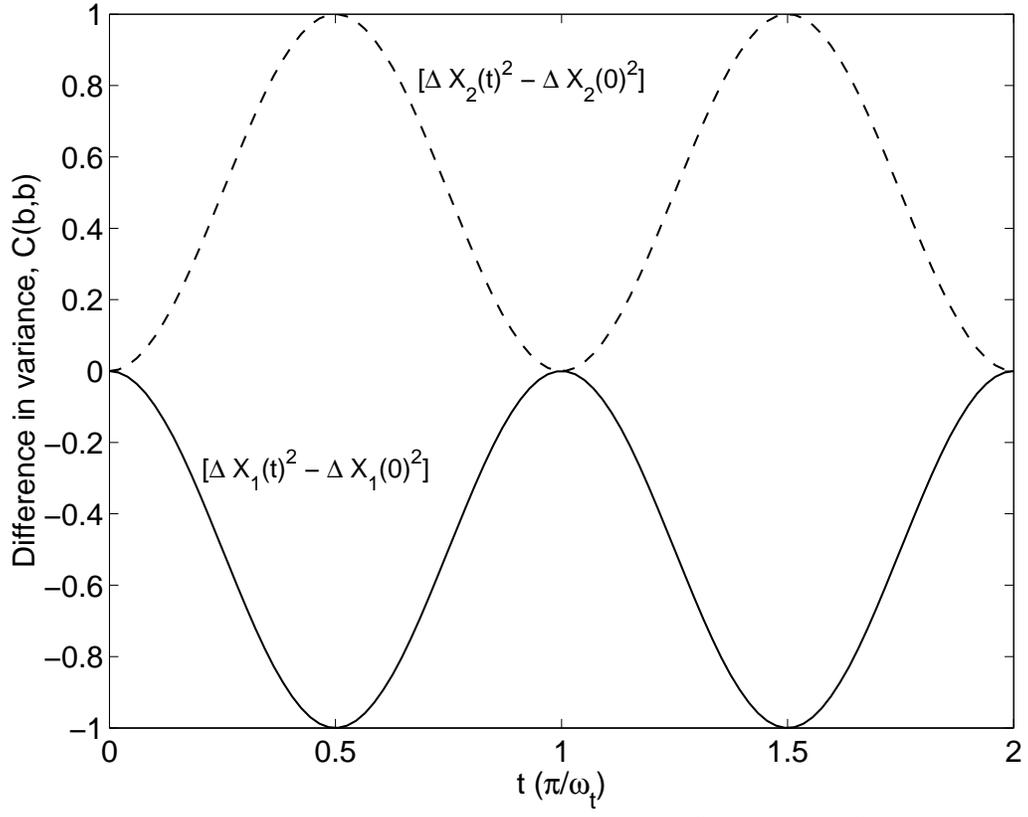}}
\caption{Change in variance of the amplitude and phase quadratures $[\Delta X_1(t)]^2$ and $[\Delta X_2(t)]^2$ as a function of time. Time is given in units of $\pi/\omega_t$ where $\omega_t$ is the trap frequency.}
\label{X1X2}
\end{figure}

It is found that the Feshbach resonance may also be used to control the
amplitude and phase quadratures of the condensate mean field. In particular, we consider
the changes induced in their respective variances i.e. the amount of quantum
mechanical squeezing\cite{walls}. Defining the usual harmonic oscillator annihilation
(creation) operator, $b$ ($b^{\dagger}$), such that $b \phi_0({\bf r}) = 0$, 
$b\phi_n({\bf r}) = \sqrt{n}\phi_{n-1}({\bf r})$ and $b^{\dagger}\phi_n({\bf %
r}) = \sqrt{n+1}\phi_{n+1}({\bf r})$, we note that, using $| \psi_c({\bf r},
t) \rangle = \sum_{n} \alpha_n(0) \exp(-i n \omega_{t} t) | \phi_n ({\bf r})
\rangle$, the number of atoms is conserved as to be expected: $\langle
\psi_c ({\bf r}, t) | b^{\dagger} b | \psi_c ({\bf r}, t) \rangle = \langle
\psi_c ({\bf r}, 0) | b^{\dagger} b | \psi_c ({\bf r}, 0) \rangle $. On the
other hand, 
\begin{equation}
\langle \psi_c ({\bf r}, t) | b | \psi_c ({\bf r}, t) \rangle = \langle
\psi_c ({\bf r}, 0) | b | \psi_c ({\bf r}, 0) \rangle e^{-i \omega_{t} t},
\end{equation}
implying that the quadrature amplitude of the field $\langle X_1 \rangle
= \frac{1}{2} \langle b + b^{\dagger} \rangle$ undergoes a time-dependent
oscillation of the form $\langle X_1(t) \rangle = \langle X_1(0) \rangle
\cos (\omega_{t} t)$ while the quadrature phase of the field is given by 
$\langle X_2(t) \rangle = \frac{1}{2i} \langle b - b^{\dagger} \rangle =
\langle X_2(0) \rangle \sin (\omega_{t} t)$. The variance $%
(\Delta X)^2 = \langle X^2 \rangle - \langle X \rangle^{2}$ of these
quadratures are given by 
\begin{eqnarray}
[\Delta X_1 (t) ]^2 & = & \frac{1}{4} \left [ C(b^{\dagger},b) +
C(b,b^{\dagger}) + C(b,b) e^{-i2\omega_{t} t} + C(b^{\dagger},b^{\dagger})
e^{i2\omega_{t} t} \right ],   \label{X1} \\
& = & \frac{1}{2} \left [ C(b^{\dagger},b) + C(b,b) \cos (2\omega_{t} t) + 
\frac{1}{2} \right ] , 
\end{eqnarray}
\begin{eqnarray}
[\Delta X_2 (t) ]^2 & = & \frac{1}{4} \left [C(b^{\dagger},b) +
C(b,b^{\dagger}) - C(b,b) e^{-i 2 \omega_{t} t} - C(b^{\dagger},b^{\dagger})
e^{i 2\omega_{t} t} \right ] , \\
& = & \frac{1}{2} \left [ C(b^{\dagger},b) - C(b,b) \cos (2\omega_{t} t ) + 
\frac{1}{2} \right ], \label{X2}
\end{eqnarray}
where we have introduced a shortened notation for the correlation between two operators $C(X,Y) = \langle XY \rangle - \langle X \rangle \langle Y \rangle$, and used
the fact that $C(b,b) \equiv C(b^{\dagger},b^{\dagger})$ for $\alpha_n$
real. In terms of the normalized expansion coefficients, 
\begin{eqnarray}
C(b,b) & \equiv & \sum_{n} \sqrt{(n+2)(n+1)} \alpha_{n}\alpha_{n+2} - \sum_{n,m} \sqrt{(n+1)(m+1)}\alpha_{n}\alpha_{n+1}\alpha_{m} \alpha_{m+1}, \\
C(b^{\dagger},b) & \equiv & \sum_{n,m} \sqrt{n}\alpha_{n}^2 - 
\sqrt{(n+1)(m+1)}\alpha_{n+1}\alpha_{n}\alpha_{m}\alpha_{m+1},
\end{eqnarray}
and since $\alpha_n \approx 0$ when $n$ is odd for the case of a harmonically trapped condensate (an even function), the correlations may also be written as $C(b,b) \approx \sum_{n} \sqrt{(2n+2)(2n+1)} \alpha_{2n}\alpha_{2n+2}$ and $C(b^{\dagger},b) \approx \sum_{n} \sqrt{2n}\alpha_{2n}^2$. Equations (\ref{X1}-\ref{X2}) indicate a time-dependent oscillation of the variances and hence quantum mechanical squeezing of these quadratures. The change from the original variances when the scattering
length was not tuned is found to be 
\begin{equation}
[\Delta X_{1} (t) ]^2 - [\Delta X_{1} (0) ]^2 = \frac{C(b,b)}{2} \left [
\cos (2\omega_ {t} t) - 1 \right ] ,
\end{equation}
\begin{equation}
[\Delta X_{2} (t) ]^2 - [\Delta X_{2} (0) ]^2 = \frac{C(b,b)}{2} \left [1 -
\cos (2 \omega_{t} t) \right ] .
\end{equation}
The maximum changes in the variances imply that the amplitude quadrature $X_1$ is squeezed periodically at the expense of the phase quadrature as a result of the change in the scattering length. The variance of the phase quadrature
is never reduced from that of the original value. Plots of these differences are presented in Fig. \ref{X1X2}. 

\begin{figure}[t]
\centerline{\psfig{height=10cm,file=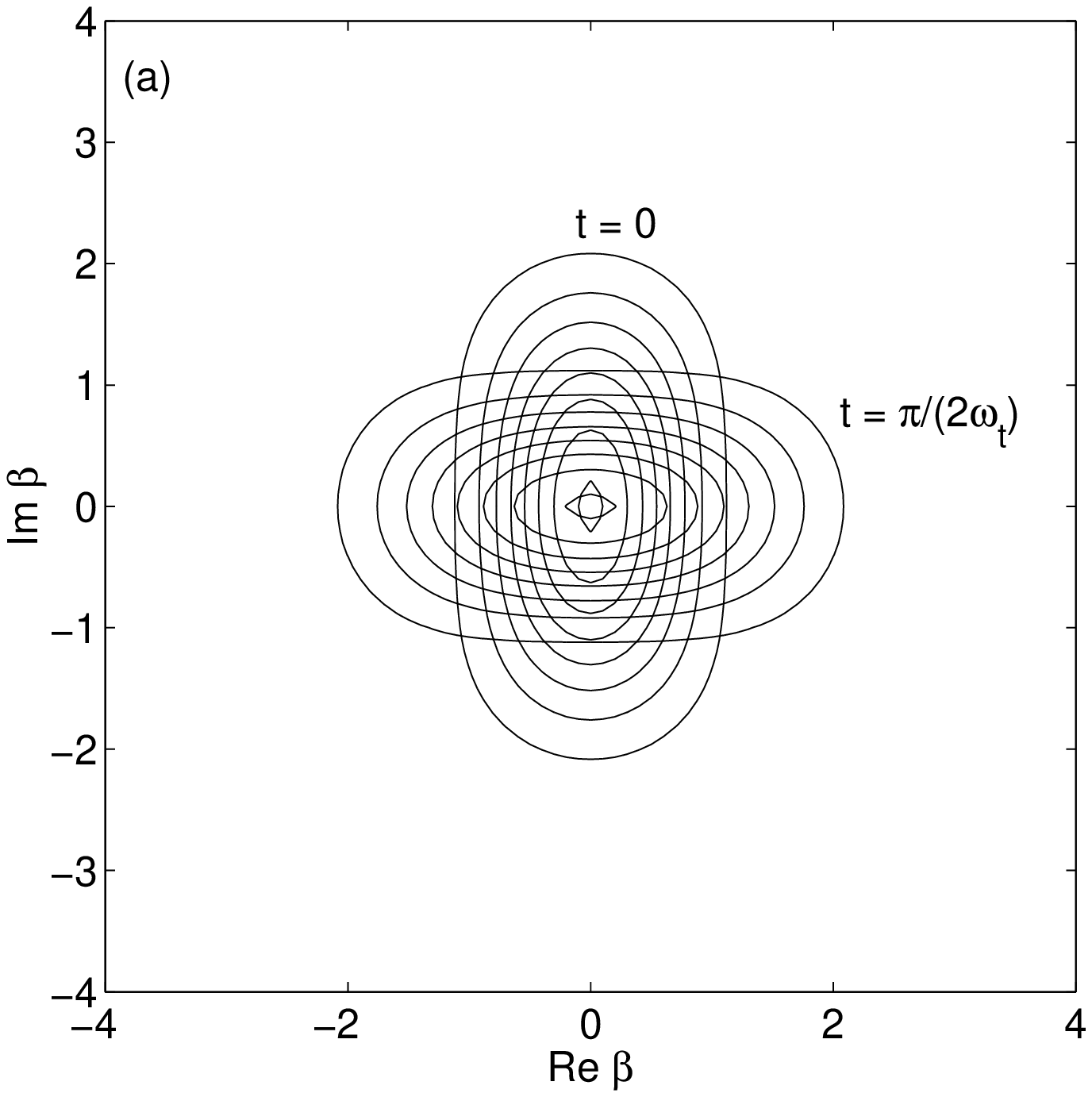} 
\psfig{height=10cm,file=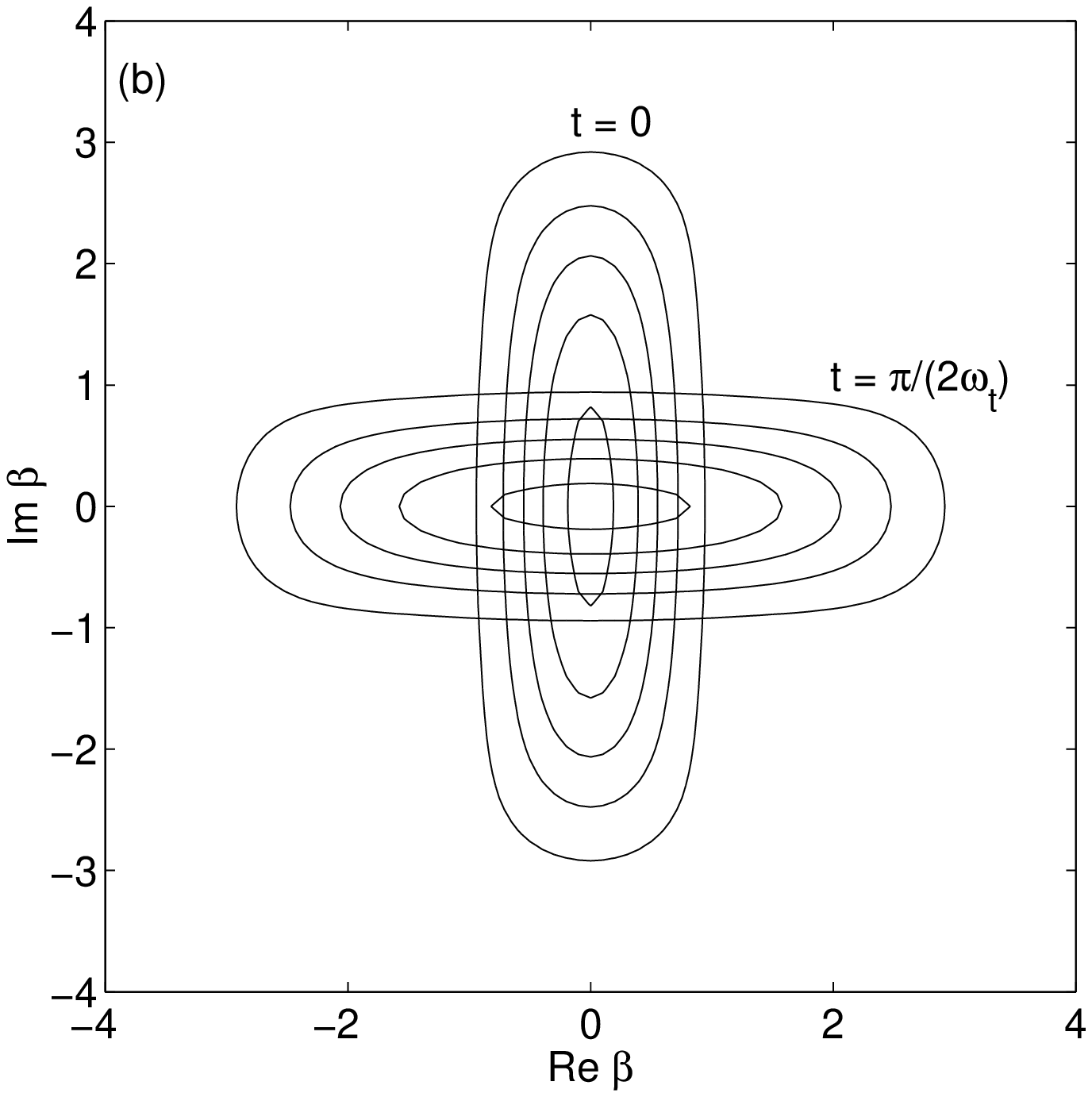}}
\caption{Wigner contour for the BEC after tuning the $s$-wave scattering length to zero at times $t = 0$ and $t = \pi/(2\omega_t)$ where $\omega_t$ is the trap frequency.
(a) for a small condensate; (b) for a BEC within the TF approximation. }
\label{Q}
\end{figure}

The induced changes in the quantum state of BEC may be visualized using the Q-function\cite{walls,Wigner} 
\begin{equation}
Q(\beta, \beta^{*}) = \langle \beta | \rho | \beta \rangle \equiv e^{-|\beta|^2}
\left | \sum_{n} \frac{\beta^{n}e^{-i\omega_{t} t} \alpha_n }{\sqrt{n!}}
\right |^2
\end{equation}
where $| \beta \rangle = e^{-|\beta|^2/2} \sum_{n} [\beta^{n}/\sqrt{n!} ] |n
\rangle$ denotes the usual coherent state with $\langle \beta | b^{\dagger}b
| \beta \rangle = |\beta|^2$ and $\rho = | \psi_c \rangle \langle \psi_c |$. 
Figure \ref{Q} shows the time development of 
the Wigner contour for the small (Gaussian) and large (TF) condensates, at times $t=0$ and $t = \pi/(2\omega_t)$. It is seen that the formerly stationary Wigner contour is now time-dependent and undergoes a rotation about its center at frequency $\omega_t$, confirming the oscillatory behavior of the quandrature squeezing. It should also be noted that, although the density variation of Fig. \ref{density} may be observed relatively straightforwardly,  the more experimentally involved quantum tomography\cite{qexp} or a projection synthesis technique\cite{qexp2} are required in order to completely measure the Q-function directly.

We note that similar idea may be applied to a system of rotating BEC. In such a case, the corresponding time-independent GPE is given by
\begin{equation}
 \left [ - \frac{\hbar^2}{2M} \nabla^{2}_{r} + \frac{1}{2}M \omega_{r}^2 r^2 + g| \psi_{rot}({\bf r})|^2  - L_z  \Omega   \right ] \psi_{rot}({\bf r}) = \mu \psi_{rot}({\bf r}),
\end{equation}
where $\omega_r$ is the frequency of harmonic potential in the $xy$ plane; ${\bf r} = (x,y)$, $L_z = |{\bf r} \times {\bf p}|$ is the angular momentum, and $\Omega$ is the rotational frequency of the trap. The eigenstates of the system after tuning the scattering length to zero are then the Landau states $\phi^{L}_{nm}({\bf r})$:
\begin{equation}
\phi^{L}_{nm}({\bf r}) = \frac{e^{|z|^2/(2a_{r}^{2})}\partial^{m}_{+} \partial^{n}_{-} e^{-|z|^2/a_{r}^2}}{\sqrt{\pi a_{r}^{2} \/ n! \/ m!}},
\end{equation}
with eigenvalues $\epsilon_{n,m} = \hbar[(\omega_{r} + \Omega)n + (\omega_{r}- \Omega)m + \omega_{r}]$, where $z = (x + iy)/a_r$, $a_r = \sqrt{\hbar / M \omega_r}$, $\partial_{\pm} =  (a_r/2)(\partial_x \pm i \partial_{y})$. It is clear that, with the nondegenerate energy levels $n' = (n,m)$, all of the previous arguments may be carried over directly to this example.

In summary, we have shown a way to execute
quantum control via Feshbach resonance -- the density and also the quantum mechanical squeezing of a trapped atomic BEC may, in principle, be controlled by tuning the $s$-wave scattering length. Experimental challenges such as the
loss of atoms\cite{Atomloss}, and the inherent noise of the external magnetic field must be
addressed in order for this scheme to be successful.

This work was supported by NSF, ONR, and ARO. 
SC wishes to thank Prof. Keith Burnett for earlier discussions.

\newpage

\end{document}